\documentclass[10pt,a4j]{article}
\usepackage[dvips]{graphicx}
\usepackage{mathrsfs}
\usepackage{eucal}
\usepackage{amsmath,amsthm,amssymb}
\usepackage{wrapfig}
\usepackage[dvips]{color}
\usepackage{cite}
\usepackage{framed}
\usepackage{tabularx}
\usepackage[sectionbib]{chapterbib}
\usepackage{chemarrow}
\usepackage{threeparttable}
\definecolor{shadecolor}{gray}{0.80}
\DeclareMathVersion{chem}
\SetSymbolFont{letters}{chem}{OT1}{cmr}{m}{n}

\begin{document}

\renewcommand{\figurename}{\small{Fig.}~}
\renewcommand{\labelitemi}{}
\renewcommand{\thefootnote}{$\dagger$\arabic{footnote}}
\renewcommand{\footnoterule}{%
  \vspace{2pt}                      
 \flushleft\rule{6.154cm}{0.4pt}   
  \vspace{4pt}                     
  }
\pagestyle{plain}

\begin{flushright}
\textit{Radius of Gyration of Branched Molecules}
\end{flushright}
\vspace{1mm}

\begin{center}
\setlength{\baselineskip}{25pt}{\LARGE\textbf{Radius of Gyration of Randomly Branched Molecules}}
\end{center}
\vspace{0mm}

\vspace*{0mm}
\begin{center}
\large{Kazumi Suematsu} \vspace*{2mm}\\
\normalsize{\setlength{\baselineskip}{12pt} 
Institute of Mathematical Science\\
Ohkadai 2-31-9, Yokkaichi, Mie 512-1216, JAPAN\\
E-Mail: suematsu@m3.cty-net.ne.jp,  Tel/Fax: +81 (0) 593 26 8052}\\[8mm]
\end{center}

\hrule
\vspace{0mm}
\begin{flushleft}
\textbf{\large Abstract}
\end{flushleft}
The mathematical derivation of the mean square radius of gyration, $\langle s_{x}^{2}\rangle$, of branched polymers is reinvestigated from a  kinetic-equation-point of view. In particular we derive the corresponding quantity of the A$-$R$-$B$_{f-1}$ model; the result showing that the mean square radius of gyration is precisely identical with that of the R$-$A$_{f}$ model.\\[-3mm]
\begin{flushleft}
\textbf{\textbf{Key Words}}:
\normalsize{Mean Radius of Gyration/ Kramers Theorem/ Kinetic Equation}\\[3mm]
\end{flushleft}
\hrule
\vspace{3mm}
\setlength{\baselineskip}{13pt}
It is well established that the mean square radius of gyration for randomly branched polymers scales as $\langle s_{x}^{2}\rangle\propto x^{\frac{1}{2}}$. This formula was derived by Zim \& Stockmayer in 1949\cite{Zim}, Dobson \& Gordon in 1964\cite{Dobson}, and Kajiwara in 1971\cite{Kajiwara}. The derivations are, however, much complicated and require harder mathematics. For this reason the result does not appear to have fully permeated into the community. In this report, we derive the same formula, in a elementary fashion, from a somewhat different point of view with the help of kinetic equation concept.

\begin{wrapfigure}[12]{r}{6cm}
\begin{center}
\vspace{-8mm}
\includegraphics[width=5cm]{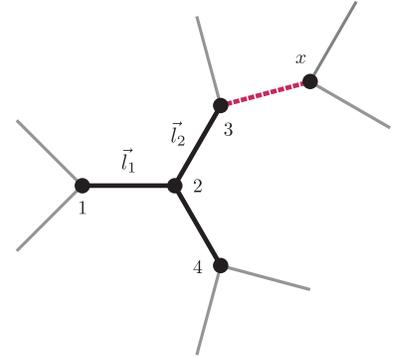}
\caption{A 3D vector field of an \textit{x}-cluster.}\label{branchedmolecule}
\end{center}
\end{wrapfigure}

\section{Kramers Theorem}
By the elementary mathematics, it is obvious that
\begin{align}
\vec{r}_{ij}=&\vec{r}_{Gj}-\vec{r}_{Gi}\\
r_{ij}^{2}=&r_{Gi}^{2}+r_{Gj}^{2}-2\hspace{0.3mm}\vec{r}_{Gi}\cdot\vec{r}_{Gj}
\end{align}
Summing over all pairs, we have
\begin{equation}
\sum_{i=1}^{N}\sum_{j=1}^{N}r_{ij}^{2}=2N\sum_{i=1}^{N}r_{Gi}^{2}-2\hspace{0.3mm}\sum_{i=1}^{N}\sum_{j=1}^{N}\vec{r}_{Gi}\cdot\vec{r}_{Gj}\label{3}
\end{equation}

\noindent By definition, $\sum_{i=1}^{x}\vec{r}_{Gi}=0$ and $s_{x}^{2}=\frac{1}{x}\sum_{i=1}^{x}r_{Gi}^{2}$. Hence, taking the statistical average of eq. (\ref{3}), we have
\begin{equation}
\left<s_{x}^{2}\right>=\frac{1}{2x^{2}}\sum_{i=1}^{x}\sum_{j=1}^{x}\left<r_{ij}^{2}\right>=\frac{1}{x^{2}}\sum_{i<j}^{x}\left<r_{ij}^{2}\right>\label{6}
\end{equation}
This formula holds whether a molecule is linear or branched. For the random flight (Brownian) chain, we have $\langle r_{ij}^{2}\rangle=l_{i}^{2}+l_{i+1}^{2}+\cdots +l_{j-1}^{2}$. Now focus our attention on any one bond of those, for instance $l_{i}$. Then it is seen that the summation in eq. (\ref{6}) simply represents the total number of trails that pass through the bond in question. 

Let all bonds have an equal length, $l$. Cutting the $k$th bond should split the \textit{x}-cluster into a \textit{k}-cluster and an $(x-k)$-cluster. Let us call these fragment clusters a $(k, x-k)$ pair. The above-mentioned number of trails is simply equal to $k\times (x-k)$. Let $\omega_{k}$ be a statistical weight to give a $(k, x-k)$ pair. There are $(x-1)\omega_{k}$ such $(k, x-k)$ pairs in the cluster. Thus we may recast eq. (\ref{6}) in the form:
\begin{equation}
\langle s_{x}^{2}\rangle=\frac{(x-1)}{x^{2}}\,l^{2}\sum_{k=1}^{x-1}\omega_{k}\, k\,(x-k)\label{7}
\end{equation}
which is just the well-known Kramers theorem\cite{Kramers}. To find the mathematical form of $\omega_{k}$, consider the equilibrium formation of an $x$-cluster, C$_{x}$, through the coupling reaction between the fragment molecules C$_{k}$ and C$_{x-k}$:
\begin{displaymath}
\text{C}_{k} \,\, +\,\, \text{C}_{x-k}\qquad \autorightleftharpoons{$K$}{}\qquad \text{C}_{x}\tag{c-1}\label{c-1}
\end{displaymath}
What we are seeking is the quantity of randomly branched polymers. So we must proceed with our calculation making use of the quantity of molecules resulting from the equilibrium reaction.
\begin{figure}[h]
\begin{center}
\includegraphics[width=8cm]{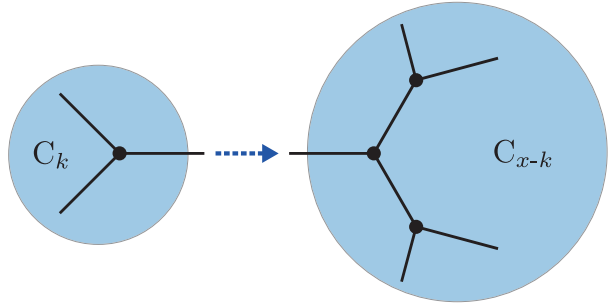}
\caption{A coupling reaction between fragment molecules.}\label{Kramers}
\end{center}
\end{figure}

\section{R$-$A$_{f}$ Model}
It has been known\cite{Stochmayer, Flory} that the population of the \textit{k}-cluster formed by eq. (\ref{c-1}) is
\begin{equation}
N_{k}=M_{0}\frac{f\{(f-1)k\}!}{k! \,\nu_{k}!}\,p^{k-1}(1-p)^{\nu_{k}}\label{8}
\end{equation}
where M$_{0}=\sum_{k=1}^{\infty}kN_{k}$ denotes the total number of monomer units and $\nu_{k}=(f-2)k+2$ the number of unreacted functional units on the \textit{k}-cluster. It is clear that there are $\nu_{k}\,\nu_{x-k}$ chances for the formation of the $k$th bond on an $x$-cluster, whereas only one chance exists for the backward (dissociation) reaction. We may drop this latter factor, since it has no effect on our final result. Hence
\begin{equation}
\omega_{k}=\frac{\nu_{k}N_{k}\cdot\nu_{x-k}N_{x-k}}{\displaystyle\sum\nolimits_{k=1}^{x-1}\nu_{k}N_{k}\cdot\nu_{x-k}N_{x-k}}\label{9}
\end{equation}
According to the mathematical theorem, the denominator reduces to
\begin{equation}
\sum_{k=1}^{x-1}\frac{\{(f-1)k\}!}{k! \,(\nu_{k}-1)!}\cdot\frac{\{(f-1)(x-k)\}!}{(x-k)! \,(\nu_{x-k}-1)!}=2(x-1)\frac{\{(f-1)x\}!}{x!\,\nu_{x}!}\label{10}
\end{equation}
Hence
\begin{equation}
\omega_{k}=\frac{x}{2}\left\{\frac{\displaystyle\frac{1}{k(x-k)}\binom{(f-1)k}{k-1}\binom{(f-1)(x-k)}{x-k-1}}{\displaystyle\binom{(f-1)x}{x-2}}\right\}\label{11}
\end{equation}
Substituting eq. (\ref{11}) into eq. (\ref{7}), we have
\begin{equation}
\langle s_{x}^{2}\rangle=\frac{(x-1)l^{2}}{2x}\,\frac{\displaystyle\sum_{k=1}^{x-1}\binom{(f-1)k}{k-1}\binom{(f-1)(x-k)}{x-k-1}}{\displaystyle\binom{(f-1)x}{x-2}}\label{12}
\end{equation}
which may be recast in the alternative form\cite{Dobson}:
\begin{equation}
\langle s_{x}^{2}\rangle=\frac{l^{2}}{2x^{2}}\,\frac{\displaystyle x! \{(f-2)x+2\}!}{\displaystyle\{(f-1)x\}!}\sum_{k=1}^{x-1}\binom{(f-1)k}{k-1}\binom{(f-1)(x-k)}{x-k-1}\label{13}
\end{equation}

\section{A$-$R$-$B$_{f-1}$ Model}
In this case, we use the number distribution\cite{Flory}:
\begin{equation}
N_{k}=M_{0}\{1-(f-1)p\}\frac{\{(f-1)k\}!}{k!\,\nu_{k}!}\,p^{k-1}\,(1-p)^{\nu_{k}}\label{14}
\end{equation}
where $\nu_{k}=(f-2)k+1$ is the number of unreacted functional units on a $k$-cluster. In this model, only A$-$B type bonds are possible, so that there are $\nu_{k}\times 1+\nu_{x-k}\times 1=\{(f-2)x+2\}$ chances for the $k$th bond formation, while a single chance exists for the dissociation reaction. Hence, dropping the constant term $\{(f-2)x+2\}$, we have
\begin{equation}
\omega_{k}=\frac{N_{k}\cdot N_{x-k}}{\displaystyle\sum\nolimits_{k=1}^{x-1}N_{k}\cdot N_{x-k}}\label{15}
\end{equation}
which, with the help of the formula (\ref{10}), again leads us to the expression:
\begin{equation}
\omega_{k}=\frac{x}{2}\left\{\frac{\displaystyle\frac{1}{k(x-k)}\binom{(f-1)k}{k-1}\binom{(f-1)(x-k)}{x-k-1}}{\displaystyle\binom{(f-1)x}{x-2}}\right\}\tag{\ref{11}$'$}
\end{equation}
Substituting into eq. (\ref{7}), we obtain
\begin{equation}
\langle s_{x}^{2}\rangle=\frac{l^{2}}{2x^{2}}\,\frac{\displaystyle x! \{(f-2)x+2\}!}{\displaystyle\{(f-1)x\}!}\sum_{k=1}^{x-1}\binom{(f-1)k}{k-1}\binom{(f-1)(x-k)}{x-k-1}\tag{\ref{13}$'$}
\end{equation}
which is exactly the same result as the foregoing solution (\ref{13}) for the R$-$A$_{f}$ model. The result is unexpected, if we recall the fact that the two systems have entirely different gelation behavior\cite{Flory}.

\section{Asymptotic Form of $\langle s^{2}_{x}\rangle$ for a Large x}
The Stirling formula is
\begin{equation}
x! \simeq\,\sqrt{2\pi x}\left(x/e\right)^{x}\label{16}
\end{equation}
for a large $x$. Let $g=f-1$ and $h=f-2$\cite{Dobson}. Applying eq. (\ref{16}) to eq. (\ref{13}), and approximating the sum by the integral, we have
\begin{align}
\hspace{0.5cm}\langle s_{x}^{2}\rangle\simeq&\,\frac{l^{2}\{hx+1\}\{hx+2\}}{2x^{2}}\sqrt{\frac{gx}{2\pi h}}\int_{1}^{x-1}\frac{\sqrt{k(x-k)}}{\{hk+1\}\{h(x-k)+1\}}dk\notag\\
=&\,\frac{l^{2}\{hx+1\}\{hx+2\}}{2x^{3/2}}\sqrt{\frac{g}{2\pi h}}\,\Bigg[\frac{2}{h^{2}}\Bigg\{\arctan \sqrt{\frac{k}{x-k}}\notag\\
&\hspace{1cm}-\frac{\sqrt{hx+1}}{hx+2}\left(\arctan \sqrt{\frac{k(1+hx)}{x-k}}+\arctan \sqrt{\frac{k}{(x-k)(1+hx)}}\right)\Bigg\}\Bigg]_{1}^{x-1}\label{17}
\end{align}
As $x\rightarrow \infty$, since $0\le |\arctan z|\le\pi/2$ by definition and $\arctan\, (\infty)=\pi/2$, the above equation reduces to
\begin{equation}
\langle s_{x}^{2}\rangle\simeq\,\left(\frac{(f-1)\pi}{2^{3}(f-2)}\right)^{1/2} x^{\frac{1}{2}}l^{2}\label{18}
\end{equation}
For a large $x$, the mean square radius of gyration of branched molecules increases as $\langle s_{x}^{2}\rangle\propto x^{1/2}$, in contrast to that of linear molecules, $\langle s_{x}^{2}\rangle\propto x$.


\end{document}